\documentclass[twocolumn,showpacs,preprintnumbers,amsmath,amssymb]{revtex4}
\usepackage{graphicx}
\usepackage{dcolumn}
\usepackage{bm}
\usepackage{epsfig}
\begin{document}
\renewcommand{\thefootnote}{\fnsymbol{footnote}}
\renewcommand{\theequation}{\arabic{section}.\arabic{equation}}

\title{Phase separation and dynamical arrest for particles interacting with mixed
potentials—the case of globular proteins revisited}

\author{Thomas Gibaud$^{1}$\footnote[1]{current address: Brandeis University, 415 South
Street, Waltham, MA-02454, USA}, Fr\'{e}d\'{e}ric Cardinaux$^{1}$, Johan Bergenholtz$^{2}$, Anna Stradner$^{1}$\footnote[2]{current address:Adolphe Merkle Institute, University of Fribourg, CH-1700 Fribourg, Switzerland}, Peter Schurtenberger$^{1}$\footnote[3]{corresponding author, current address: Division of Physical Chemistry, Center for Chemistry and Chemical Engineering, Lund University, SE-221 00 Lund, Sweden.}}
  \affiliation{(1) Department of Physics and Fribourg Center for Nanomaterials, University of Fribourg, CH-1700 Fribourg, Switzerland}
   \affiliation{(2) Department of Chemistry, G\"oteborg University, SE-412 96, G\"oteborg, Sweden}

\begin{abstract}
We examine the applicability of the extended law of corresponding
states (ELCS) to equilibrium and non equilibrium features of the
state diagram of the globular protein lysozyme. We provide
compelling evidence that the ELCS correctly reproduces the location
of the binodal for different ionic strengths, but fails in describing
the location of the arrest line. We subsequently use Mode Coupling
Theory (MCT) to gain additional insight into the origin of these
observations. We demonstrate that while the critical point and the
connected binodal and spinodal are governed by the integral features
of the interaction potential described by the normalized second virial
coefficient, the arrest line is mainly determined by the attractive well
depth or bond strength. This article is published in Soft Matter. The reference is: DOI: 10.1039/c0sm01175d
\end{abstract}
\maketitle
\section{Introduction}
Gel formation in colloidal suspensions has continued to attract
considerable attention during the last few years. Particular attention
has been given to particles interacting via a short-ranged attractive
potential, and the role of the key parameters volume fraction $\phi$,
interparticle interaction strength Ua and range of the potential D in
determining the state diagram has been addressed [1,2]. Depending upon
the strength of the attraction, they can form amorphous solids from
arbitrarily low volume fraction, $\phi$, up to random close packing.1 At
intermediate values of $U_{a}/kT$ and $\phi$, one encounters an intriguing
interplay between spinodal decomposition and gelation that results in
the phenomenon of an arrested spinodal decomposition and the
formation of a solid-like network [2–6]. While there are a number of
experimental studies that report on the phenomena, there are relatively
few quantitative results on the actual location of the arrest line
in the vicinity of the binodal and spinodal. In a recent study Lu et al.6
provided for example evidence that gelation for short range attractive
particles indeed occurs as a consequence of an initial equilibrium
liquid–liquid phase separation. The authors claimed that all phase
separating samples gelled, i.e., that the gel line coincided with the
binodal in the Baxter sticky sphere model, and that the concentration
of the resulting colloidal gels would always be around $\phi \sim$ 0.55. This
was interpreted as a result of the gel line intersecting the phase
separation boundary at $\phi \sim$ 0.55, with the dense phase undergoing
an attractive glass transition after phase separation.

Moreover, these authors also suggested the existence of a universal
state diagram when using the second virial coefficient $B_{2}$, normalized
to the second virial coefficient $B_{HS}=2\pi\sigma^{3}/3$ of hard spheres with the
same diameter $\sigma$ , as an effective temperature $B_{2}^{*}=B_{2}/B^{2}_{HS}$. It had in
fact already previously been suggested that the so-called extended law
of corresponding states (ELCS) [7] should not only allow for a rescaling
of thermodynamic properties, but that it should also hold for
dynamic properties such as gel or glass lines that describe the
occurrence of dynamical arrest [2,8].

Another example, where a state diagram that combines liquid–
liquid phase separation and dynamical arrest has been reported, was
based on the use of globular proteins as model colloids [4,9]. Direct
analogies between colloids interacting via a short range attractive
potential and globular proteins have been used in the past quite
successfully to understand the phase behavior of globular proteins
and to rationalize some of the phenomenological observations made
during protein crystallization [10$–$14]. In particular lysozyme has been
extensively investigated in view of the potential but also limits of such
a coarse-grained colloid approach to protein solutions [11,13,15,16].

While the existence of dynamical arrest in protein solutions has in
fact already been reported previously, [13] it is only recently that we have
been able to locate the glass line in the region below the coexistence
curve for the globular protein lysozyme [4]. These findings are
summarized in the lysozyme state diagram in Fig. 1, where the
temperature $T$ is plotted as a function of the protein volume fraction
$\phi$. A lysozyme solution above the coexistence curve quenched down
to temperatures within region I undergoes complete macroscopic
phase separation after some time [4]. However, deeper quenches into
region II lead to the formation of a gel due to the interplay between
phase separation and dynamical arrest [4,9].

Arrest is thought to occur during the early stage of spinodal
decomposition when the volume fraction $\phi_{2}$ of the dense phase
intersects the dynamical arrest threshold $\phi_{2,Glass}$ given by the arrest
line, which extends deep into the unstable region (right triangles in
Fig. 1). At this point the dense phase forms an attractive glass and
arrests, and phase separation thus gets pinned into a space spanning
gel network with a characteristic length $\xi$ [4]. Additional support for this
scenario was subsequently obtained through investigations
combining video microscopy and ultra small-angle light scattering
experiments [9]. Moreover, this interpretation is also in agreement with
the work of Dumetz et al.,[16] who looked at the phase behavior of
several globular proteins and noted the existence of an arrested state.

\begin{figure}[h]
    \centering
    \includegraphics[width=5cm]{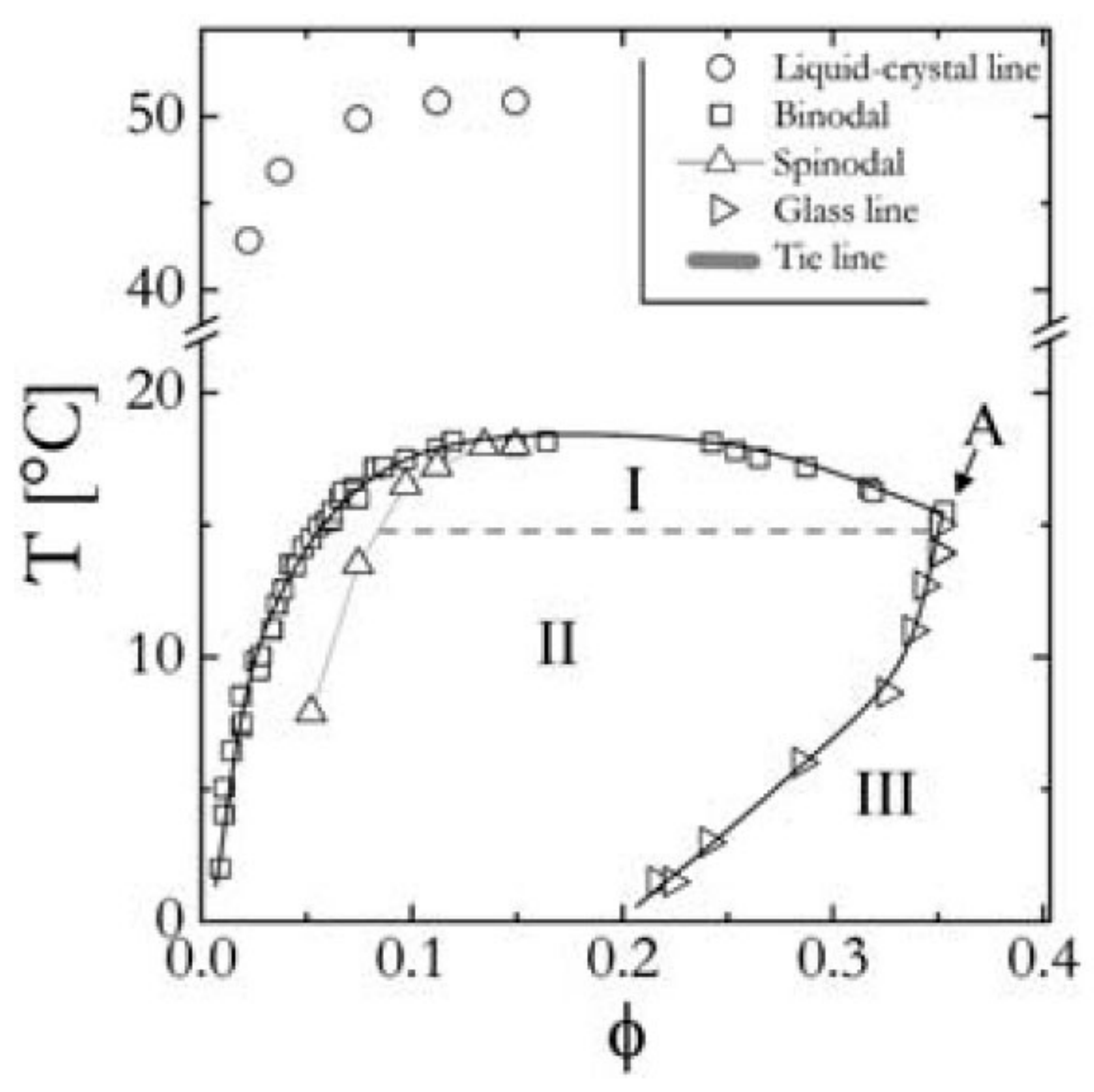}
        \caption{State diagram of lysozyme suspended in an aqueous Hepes buffer
at pH of 7.8, with [NaCl] = 500mM. Shown are fluid–crystal coexistence
curve (circles), binodal (squares), spinodal (triangles) and dynamical
arrest line (right triangles), and the tie line separating region I from II. In
region I we observe phase separation, in region II arrested spinodal
decomposition, and region III corresponds to the attractive glass (see ref.
4 for details).}
    \label{Fig1}
    \end{figure}

If the invariance of the state diagram for colloids with short range
attractions in the frame of a Noro-Frenkel ELCS would also hold for
non-equilibrium properties such as gel and glass formation as
proposed by a number of authors, [6,8,17] this could serve as an ideal basis
for comparing various systems and generating a generic phase diagram
as a function of a single control parameter $B_{2}^{*}$. However, while there
exist simulations that have addressed this problem, [8] there are to our
knowledge no systematic experimental investigations available that
have tested the applicability of the ELCS to glass or gel formation for
colloidal suspensions. Moreover, while most theoretical, simulation
and experimental studies have concentrated on purely attractive
systems, any attempt to extend this approach to protein solutions and
many colloidal systems of practical importance will have to consider
the effects of amixed potential that also includes a soft repulsion due to
residual charges.While it has already been observed that the location
of the binodal changes with ionic strength in a way that is at least
qualitatively in agreement with the ELCS, [10] no information on the
arrest line in the vicinity of the binodal has been reported so far, despite
the fact that this is for example an essential ingredient of the zone
picture for protein crystallization. We have thus started a systematic
characterization of the binodal and the arrest line under conditions
where we increase the contributions from the screened Coulomb
repulsion. We in particular concentrate on the question whether the
resulting liquid–gas coexistence curves and the arrest lines can be
superimposed by using $B_{2}^{*}$ as a scaling parameter also if the electrostatic
repulsion becomes more prominent and longer ranged.

\section{Material and Methods}
We used the globular protein hen egg white lysozyme (Fluka,
L7651) in 20mM HEPES buffer at pH = 7.8, where lysozyme carries
a net charge of +8e.18 We varied the electrostatic repulsion created by
the residual charges of lysozyme at pH=7.8 by adding different
amounts of NaCl. Details about the sample preparation are given
elsewhere [4]. The buffers used in this study contain 200, 300, 400, and
500mM NaCl. Taking into account the counter-ions from the
proteins, this corresponds to a decrease of the Debye length $l_{D}$ from
about 19\% of the lysozyme diameter for 200mM to about 12\% for
500mM NaCl for a lysozyme solution at $c=$20 mg mL$^{-1}$.19 For the
determination of the state diagrams at different ionic strengths we
again used a combination of visual observation, optical microscopy,
static light scattering, rheology and centrifugation as described in
ref.~4. The volume fraction of lysozyme f was calculated from the
protein concentrations c as measured by UV absorption spectroscopy
using $\phi=c/\rho$, where $\rho= 1.351$ g cm$^{3}$ is the protein density.

The second virial coefficient for lysozyme solutions was measured using static light scattering experiments on concentration series of dilute solutions of lysozyme. The light scattering experiments were performed with a commercial goniometer system
(ALV/DLS/SLS-5000F mono-mode fiber compact goniometer system with ALV-5000 fast correlator) at a wavelength of 514.5nm and a fixed scattering angle of 90$^{o}$, leading to a scattering vector $q$ = 0,022 nm.
Figure 2 summarizes the results from our determination of the second virial coefficient B2 as a function of temperature for four different values of the ionic strength. Here we used the relation $Kc/R = 1/M+ 2N_{A}B_{2}c/M^{2}$, where $K$ is the contrast term, $c$ is the
concentration, $M$ is the molecular weight, and $N_{A}$ is the Avogadro's number, to determine $B_{2}$ from the measured Rayleigh ratios $R$. Plotted in Fig. 2 are the values of the second virial coefficients obtained and normalized by their value for hard sphere, $B^{HS}_{2} = 2\pi\sigma^{3}/3$. Also shown are the values of the critical temperature $T_{c}$ for the different salt concentrations. It is interesting to note that for all salt concentrations investigated, the $B_{2}^{*}=B2/B^{HS}_{2}$ values at $T_{c}$ are $B_{2}^{*}=2.55\pm0.12$

\begin{figure}[h]
    \centering
    \includegraphics[width=5cm]{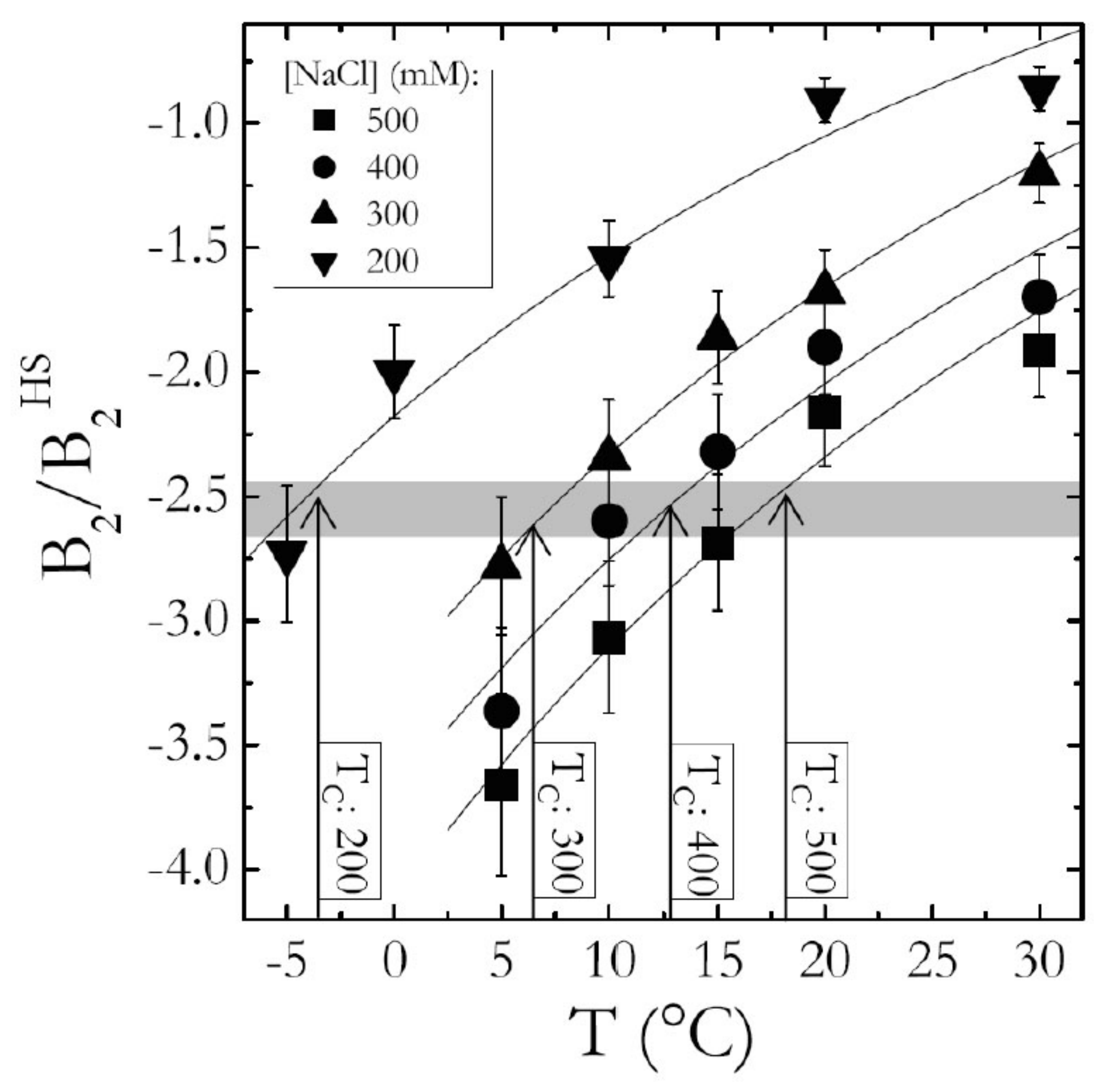}
        \caption{Temperature dependence of the normalized second virial coefficient $B2/B^{HS}_{2}$ obtained from light scattering experiments for different
ionic strengths. Lines are guides to the eye. Vertical dotted lines highlight the value of the critical temperature $T_{c}$ for a given value of the ionic
strength.}
    \label{Fig2}
    \end{figure}

\section{Results}
We first determined the location of the coexistence curves as
a function of salt concentration. The resulting binodals are shown in
Fig. 3A.We find that the critical volume fraction $\phi_{c}=0.17 \pm 0.02$ is
salt independent and consistent with values from literature [10,20]. The
critical temperature $T_{c}$, however, increases with increasing salt
concentration, reflecting the fact that the attractive part of the interaction
potential which drives phase separation becomes more dominant
as the electrostatic repulsion becomes increasingly screened.

It has already been demonstrated in various studies of liquid–gas
coexistence and crystallization in protein solutions that the ionic
strength dependence of the location of the liquid–solid and liquid–
liquid phase boundaries can be rationalized at least qualitatively using
an integral quantity of the strength of the interaction potential as
given by $B_{2}$ [11,13,16,21$–$25]. We have thus also determined the second virial
coefficient $B_{2}$ as a function of temperature for all ionic strengths with
static light scattering (see ESI† for additional information).We found
that the normalized $B_{2}^{*}$ at $T_{c}$ is $B_{2}^{*}=2.55\pm0.12$, identical within
experimental errors irrespective of ionic strength, in qualitative
agreement with the idea that equilibrium properties are dictated by
integral features of the interaction potential.We can in fact rescale the
corresponding phase boundaries for all values of the ionic strength
using $B_{2}^{*}$ as an effective temperature as shown in Fig. 3B. This
provides us with a direct verification of the applicability of the ELCS
for lysozyme as a model for particles interacting via amixed potential.

In a next step we now focus on the dynamical arrest line as
a function of ionic strength and temperature. We determined their
location by a combination of rheology, microscopy and centrifugation
experiments as described in detail in ref. 4. The results are also
summarized in Fig. 4. In contrast to the strongly salt-dependent
coexistence curves, the location of the dynamical arrest line seems to
be independent of ionic strength, and all the arrest lines at different
ionic strengths overlap within the experimental error bars. This holds
not only for the points within the unstable region that were determined
by a combination of a rapid temperature quench and subsequent
centrifugation, but also for the intersection with the binodal
(point A in Fig. 1) determined independently by rheology and
microscopy.

\begin{figure}[h]
    \centering
    \includegraphics[width=8cm]{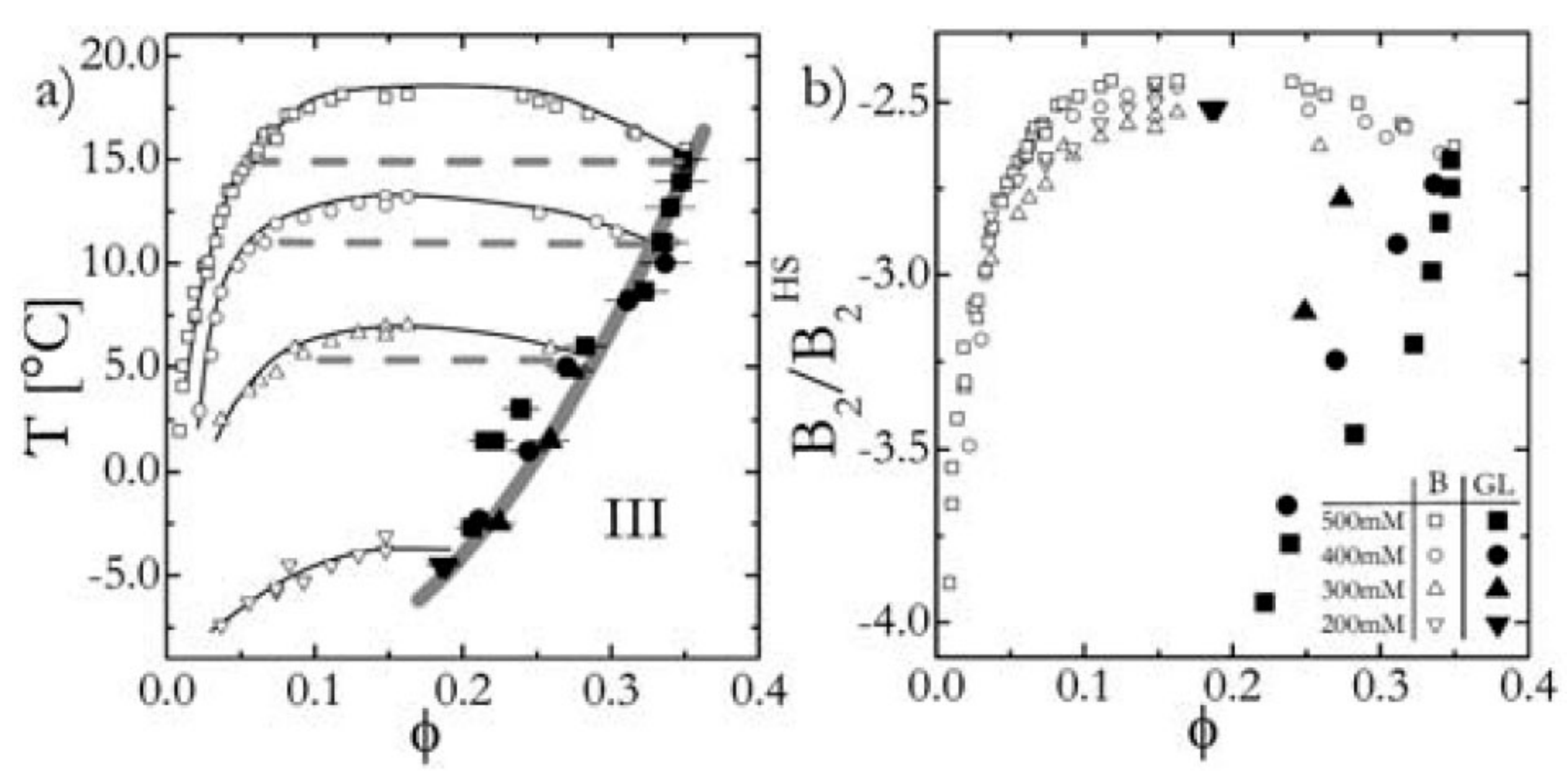}
        \caption{Lysozyme state diagram showing the metastable binodal (open
symbols) and the dynamical arrest lines (full symbols) for different ionic
strengths. (a) $T$ vs. $\phi$ presentation: the dynamical arrest lines for different
values of the ionic strength all overlap. Lines are guides to the eye. (b) $T_{eff}$
vs. $\phi$ presentation: in agreement with the ELCS, the binodal lines all
overlap.}
    \label{Fig2}
    \end{figure}

These experimental observations thus lead to the following
scenario: the coexistence curves are not sensitive to the details of the
interaction potential and can be rescaled using an integral quantity—
the second virial coefficient—that characterizes the overall potential.
However, this drastically changes when it comes to dynamical arrest.
Here it appears that the short-range part of the potential, i.e. the
temperature-dependent attractive well, determines its location.

We next turn to the idealized mode coupling theory (MCT) [27] in an
attempt to gain some qualitative insight into the trends observed
experimentally. MCT requires the static structure factor $S(q)$ as input,
which we obtain from Percus–Yevick (PY) theory [28] applied to an
interaction potential of double Yukawa form,
$U(r)/kT=-U_{0}/kT[K_{1}e^{-z_{1}(r-1)}+K_{2}e^{-z_{2}(r-1)}]/[(K_{1}+K_{2})r]$
for separation distances $r > 1$ and complemented by a hard core for $r
< 1$. Here we chose to use a simple mixed potential that would allow
us to capture the general trends upon a variation of the ionic
strengths. The parameters of the screened Coulomb part of the
interaction, $K_{2} > 0$ and $z_{2}$, have been calculated from the experimental
parameters using the linear superposition principle of the
linearized Poisson–Boltzmann equation. Their (weak) temperature
dependence has been neglected. In addition, $K_{1} < 0$ and $z_{1}=50$, so
that the first term in the interaction potential models a short-range
attraction and the contact value provides a suitable 'temperature', \emph{viz.}
$-kT/U_{0}$, for a qualitative comparison with the experimental data of
Fig. 3. However, we refrain from attempting to model the nonequilibrium
transitions quantitatively because of the limitations of
MCT for very short-range attractions.8

Fig. 4A shows the arrest line observed within MCT as a function of
added monovalent salt. At the higher volume fractions increasing the
ionic strength has the effect of shifting the glass transitions to higher
$\phi$, in agreement with ref. 26. However, at sufficiently low temperatures
and particle concentrations the glass lines merge and become
essentially independent of the ionic strength. In this region the location
of the glassy arrest transition produced by the theory is found to
be governed by the contact value of the interaction potential. Indeed,
at comparable f the location of the glass line is unaffected by changes
in the ionic strength (see Fig. 4A). MCT thus captures the experimentally
observed trends. This is further demonstrated in Fig. 4B,
where the theoretical predictions are scaled by $B_{2}^{*}$. Whereas PY
calculations for the spinodals reveal that they superpose reasonably
well under these conditions (data not shown), the glass lines from the
theory do not scale with $B_{2}^{*}$, confirming that under the prevailing
conditions it is the well depth or 'bond strength', rather than $B_{2}^{*}$, that
governs the structural arrest.

\begin{figure}[h]
    \centering
    \includegraphics[width=8cm]{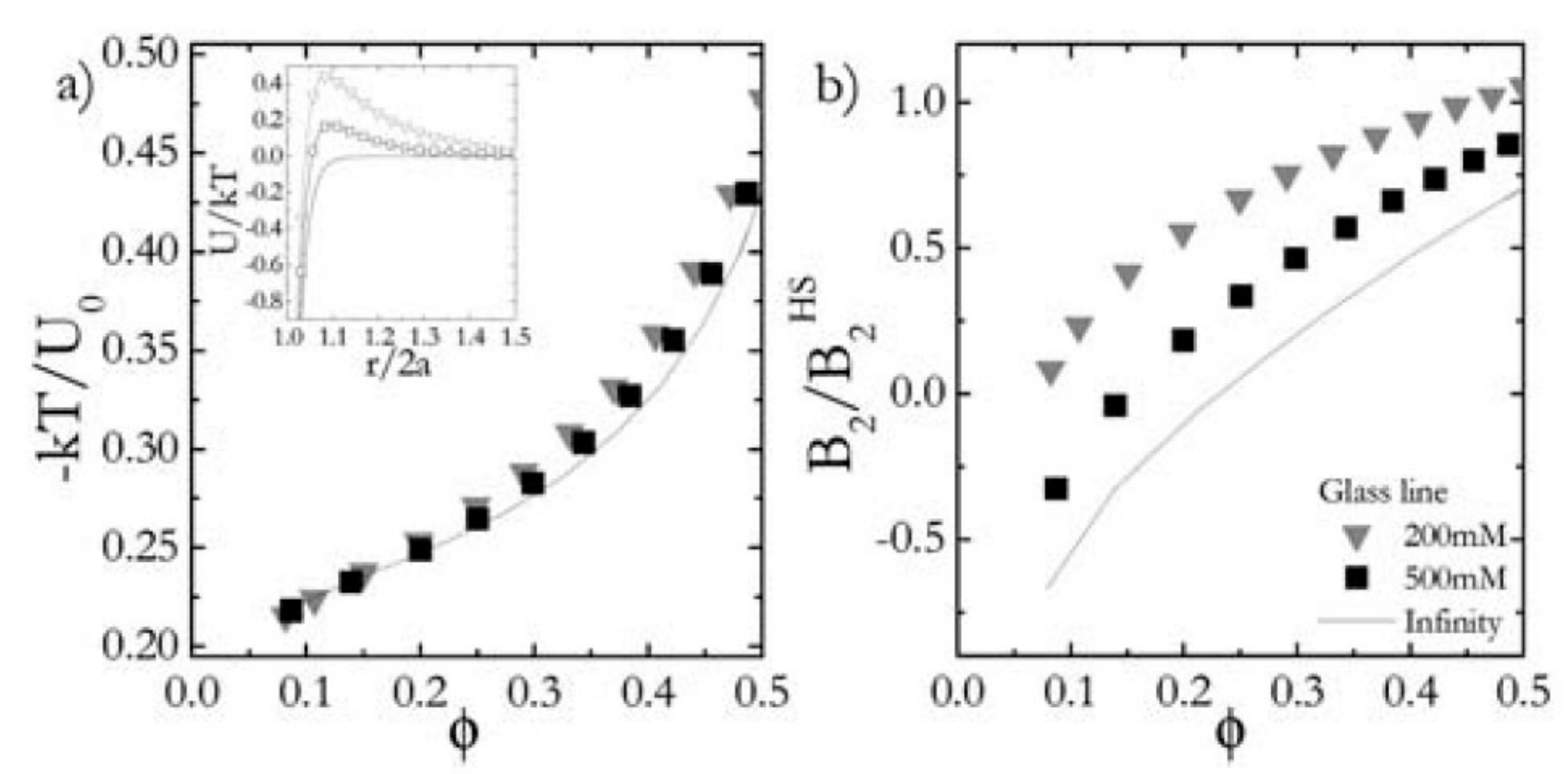}
        \caption{Attractive glass transition determined from MCT for [NaCl]=$\infty$
(pure attraction), 500 mM, and 200mM NaCl. Plotted is the inverse
contact value of the interaction potential (a) and the reduced second virial
coefficient (b) as a function of the particle volume fraction. Calculations
have been made on the basis of a double Yukawa interaction, as
described in ref. 26, some examples of which are shown in the inset for
different ionic strengths.}
    \label{Fig3}
    \end{figure}

We are thus left with the discrepancy between the results from ref.
8, where evidence was presented that the ELCS would also be valid
for dynamical arrest, and our own findings. However, it is important
to point out that the simulations performed in ref. 8 were performed
for purely attractive potentials of very short range $\varepsilon=\Delta/a\leq 0.01$.On
the other hand, lysozyme possesses a mixed potential, and has been
modeled with a range for the attraction of approximately $0.04\leq\varepsilon\leq0.34$, depending upon whether an isotropic potential or a patchy
description has been used [15]. It is clear thatMCT provides qualitative
agreement between experiment and theory only. However, despite its
shortcomings,MCT should be able to reproduce dynamical arrest for
particles interacting via a not too short-ranged potential reasonably
well, [29] and indeed captures the trends observed (Fig. 3 and 4).

\section{Conclusion}
Additional simulations and experiments with colloidal model
systems under comparable conditions (mixed potential, not too
short-ranged attraction) are thus needed before we can generalize our
findings. However, we believe that our results are of considerable
importance for current attempts to link proteins and colloids. A
coarse-grained approach to the phase behavior of globular proteins
based on a close analogy between colloids and proteins has frequently
been used to gain insight into phenomena such as protein crystallization
or protein condensation diseases [10,13,14,30]. The assumption of
a generalized state diagram has for example been used to develop new
strategies in protein crystallization. However, our study clearly
demonstrates that there may be no sound basis for such a generic
state diagram for globular proteins that incorporates both equilibrium
as well as non-equilibrium features.

We may also speculate about the importance of our findings for the
colloid community. For colloids with mixed interaction potentials we
may observe a significant shift of the arrest line with respect to the
binodal at different ionic strengths. For weakly screened electrostatic
repulsion this could result in a shift of the binodal below the arrest line
and thus open up a direct path to gelation that does not compete with
a simultaneous spinodal decomposition. This is a scenario that has
been discussed previously in attempts to generalize our current
understanding of gelation and glass formation, but no convincing
experimental evidence has been provided so far.2Given the enormous
structural and mechanical differences found in colloidal gels and
glasses that have formed via different routes, this opens up new
avenues towards the formation of solid-like complex fluids that may
be extremely interesting for applications in materials or food science [31].

This work was supported by the Swiss National Science Foundation,
the State Secretariat for Education and Research (SER) of
Switzerland, the Marie Curie Network on Dynamical Arrest of Soft
Matter and Colloids (MRTN-CT-2003-504712), and the EU NoE
SoftComp NMP3-CT-2004-502235. We also acknowledge support
from the Adolphe Merkle Foundation.

\section{Bibliography}
\begin{enumerate}
  \item V. Trappe and P. Sandkuhler, \emph{Curr. Opin. Colloid Interface Sci.}, 2005,
\textbf{18}, 494–500.
  \item E. Zaccarelli, \emph{J. Phys.: Condens. Matter}, 2007, \textbf{19}, 323101.
  \item  N. A. M. Verhaegh, D. Asnaghi, H. N. W. Lekkerkerker, M. Giglio
and L. Cipelletti, \emph{Phys. A}, 1997, \textbf{242}, 104–118.
  \item  F. Cardinaux, T. Gibaud, A. Stradner and P. Schurtenberger, \emph{Phys.
Rev. Lett.}, 2007, \textbf{99}, 118301.
  \item  S. Buzzaccaro, R. Rusconi and R. Piazza,\emph{ Phys. Rev. Lett.}, 2007, \textbf{99},
098301.
  \item  P. Lu, E. Zaccarelli, F. Ciulla, A. B. Schofield, F. Sciortino and
D. A. Weitz,\emph{ Nature}, 2008, \textbf{453}, 499–503.
  \item  M. G. Noro and D. Frenkel, \emph{J. Chem. Phys.}, 2000, \textbf{113}, 2941–2944.
  \item  G. Foffi, C. de Michele, F. Sciortino and P. Tartaglia, \emph{Phys. Rev.
Lett.}, 2005, \textbf{94}, 078301.
  \item  T. Gibaud and P. Schurtenberger, \emph{J. Phys.: Condens. Matter}, 2009,
\textbf{21}, 322201.
  \item  M. Muschol and F. Rosenberger, \emph{J. Chem. Phys.}, 1997, \textbf{107}, 1953.
  \item  R. Piazza, V. Peyre and V. Degiorgio, \emph{Phys. Rev. E: Stat. Phys.,
Plasmas, Fluids, Relat. Interdiscip. Top.}, 1998, \textbf{58}, R2733.
  \item  R. Piazza, Curr. Opin. Colloid Interface Sci., 2000, \textbf{5}, 38.
  \item  A. Kulkarni, N. M. Dixit and C. F. Zukoski, \emph{Faraday Discuss.}, 2003,
\textbf{123}, 37.
  \item  G. Foffi, K. A. Dawson, S. V. Buldyrev, F. Sciortino, E. Zaccarelli
and P. Tartaglia, \emph{Phys. Rev. E: Stat. Phys., Plasmas, Fluids, Relat.
Interdiscip. Top.}, 2002, \textbf{65}, 050802.
  \item  C. Gogelein, G. Nagele, R. Tuinier, T. Gibaud, A. Stradner and
P. Schurtenberger, \emph{J. Chem. Phys.}, 2008, \textbf{129}, 085102.
  \item  A. Dumetz, A. M. Chockla, E. W. Kaler and A. M. Lenhoff, \emph{Biophys.
J.}, 2008, \textbf{94}, 570.
  \item  E. Zaccarelli, P. J. Lu, F. Ciulla, D. A. Weitz and F. Sciortino, J.
Phys.: Condens. Matter, 2008, \textbf{20}, 494242.
  \item  C. Tanford and R. Roxby, Biochemistry, 1972, \textbf{11}, 2192.
  \item  L. Belloni, \emph{J. Chem. Phys.}, 1986, \textbf{85}, 519–526.
  \item  V. G. Taratuta, A. Holschbach, G. M. Thurston, D. Blankschtein and
G. B. Benedek, \emph{J. Phys. Chem.}, 1990, \textbf{94}, 2140–2144.
  \item  A. George and W. W. Wilson, \emph{Acta Cryst.}, 1994, 50, 361–365.
  \item  D. F. Rosenbaum and C. F. Zukoski, \emph{J. Cryst. Growth}, 1996, \textbf{169},
752.
  \item  M. L. Broide, T. M. Tominc and M. D. Saxowsky, \emph{Phys. Rev. E:
Stat. Phys., Plasmas, Fluids, Relat. Interdiscip. Top.}, 1996, \textbf{53},
6325.
  \item  D. F. Rosenbaum, A. Kulkarni, S. Ramakrishnan and C. F. Zukoski,
\emph{J. Chem. Phys.}, 1999, \textbf{111}, 9882.
  \item  H. Sedgwick, J. E. Cameron, W. C. K. Poon and S. U. Egelhaaf, J.
Chem. Phys., 2007, \textbf{127}, 125102.
  \item  J. Wu, Y. Liu, W.-R. Chen, J. Cao and S.-H. Chen, \emph{Phys. Rev. E:
Stat., Nonlinear,Soft Matter Phys.}, 2004, \textbf{70}, 050401.
  \item  W. Gotze and L. Sjogren, \emph{Rep. Prog. Phys.}, 1992, \textbf{55}, 241.
  \item  B. D'Aguanno and R. Klein, Static scattering properties of colloidal
suspensions, Clarendon Press, Oxford, 1996.
  \item  B. Ahlstr€om and J. Bergenholtz,\emph{ J. Phys.: Condens. Matter}, 2007, \textbf{19},
036102.
  \item  A. Stradner, G. Foffi, N. Dorsaz, G. Thurston and P. Schurtenberger,
\emph{Phys. Rev. Lett.}, 2007, \textbf{99}, 198103.
  \item  R. Mezzenga, P. Schurtenberger, A. Burdige and M. Michel, \emph{Nat.
Mater.}, 2005, \textbf{4}, 729.

\end{enumerate}

\end{document}